\documentclass[acmlarge]{acmart}
\makeatletter
\newcommand{\confshort}{\acmConference@shortname}
\newcommand{\conffull}{\acmConference@name}
\newcommand{\confdate}{\acmConference@date}
\newcommand{\confloc}{\acmConference@venue}
\AtBeginDocument{
  \fancypagestyle{firstpagestyle}{
    \fancyhead{}%
    \fancyfoot[C]{}%
  }
  \fancyhf{}
  \fancyhead[LO]{\@headfootfont\shorttitle}%
  \fancyhead[RE]{\@headfootfont\@shortauthors}%
  \fancyhead[LE]{\@headfootfont\footnotesize \confshort, \confdate, \confloc}%
  \fancyhead[RO]{\@headfootfont\footnotesize \confshort, \confdate, \confloc}%
  \fancyfoot[C]{}%
}
\makeatother
\acmBooktitle{\conffull\@ (\confshort), \confdate, \confloc}

\AtBeginDocument{%
  }

\copyrightyear{2026}
\acmYear{2026}
\setcopyright{cc}
\setcctype{by}
\acmConference[FAccT '26]{The 2026 ACM Conference on Fairness, Accountability, and Transparency}{June 25--28, 2026}{Montreal, QC, Canada}
\acmBooktitle{The 2026 ACM Conference on Fairness, Accountability, and Transparency (FAccT '26), June 25--28, 2026, Montreal, QC, Canada}
\acmDOI{10.1145/3805689.3812293}
\acmISBN{979-8-4007-2596-8/2026/06}
\usepackage{amsmath}
\usepackage{amsthm}
\newtheorem{definition}{Definition}[section]

\begin{document}

\title{The Paradox of Prioritization in Public Sector Algorithms}

\author{Erina Seh-Young Moon}
\affiliation{%
\institution{University of Toronto}
  \city{Toronto}
  \country{Canada}}
\email{erina.moon@mail.utoronto.ca}
\orcid{0000-0003-3233-9773}

\author{Shion Guha}
\affiliation{%
\institution{University of Toronto}
  \city{Toronto}
  \country{Canada}
 }
\email{shion.guha@utoronto.ca}
\orcid{0000-0003-0073-2378}

\renewcommand{\shortauthors}{Moon et al.}

\begin{abstract}
Public sector agencies perform the critical task of implementing the redistributive role of the State by acting as the leading provider of critical public services that many rely on. In recent years, public agencies have been increasingly adopting algorithmic prioritization tools to determine which individuals should be allocated scarce public resources. Prior work on these tools has largely focused on assessing and improving their fairness, accuracy, and validity. However, what remains understudied is how the structural design of prioritization itself shapes both the effectiveness of these tools and the experiences of those subject to them under realistic public sector conditions. In this study, we demonstrate the fallibility of adopting a prioritization approach in the public sector by showing how the underlying mechanisms of prioritization generate significant relative disparities between groups of intersectional identities as resources become increasingly scarce. We argue that despite prevailing arguments that prioritization of resources can lead to efficient allocation outcomes, prioritization can intensify perceptions of inequality for impacted individuals. We contend that efficiencies generated by algorithmic tools should not be conflated with the dominant rhetoric that efficiency necessarily entails “doing more with less” and we highlight the risks of overlooking resource constraints present in real-world implementation contexts.
\end{abstract}

\ccsdesc[500]{Human-centered computing~Empirical studies in HCI}
\ccsdesc[500]{Applied computing~Computing in government}

\keywords{prioritization, algorithm, public sector, resource allocation, government}

\maketitle

\section{Introduction}

Public sector agencies perform the critical task of carrying out the redistributive role of the state and have increasingly adopted algorithmic prioritization tools to allocate essential resources effectively and efficiently \cite{Reutter_and_2024, showkat23, kelly23, moon24, johnson22, eubanks2018automating}. For example, in North America, homelessness service systems increasingly rely on predictive risk algorithms to assess the vulnerability of clients seeking housing supports in order to prioritize services to them, and public schools employ tools to determine which students should be prioritized to receive academic support to prevent school dropout \cite{eubanks2018automating, showkat23, wang2024, kelly23, perdomo2023}. Public sector prioritization algorithms are distinct from private sector decision-making algorithms as the tools are often applied to the most vulnerable members of society and states take an active role in controlling these populations \cite{levy21, eubanks2018automating, bridges17}. 

Existing scholarship on public sector algorithms has largely focused on improving the models themselves, with an emphasis on enhancing model accuracy, fairness, and validity \cite{sambasivan21, liu2024, mothilal24}. Yet, there have been growing calls within the algorithmic fairness community to move beyond this model-centered approach and the need to examine these tools as policy interventions in practice while considering their implementation context \cite{shirali24, liu2025bridging, liu2024}. In this work, we respond to these calls by shifting the focus from the algorithmic model to the very structural mechanisms of prioritization that underpins public sector prioritization algorithms. We define structural mechanisms of prioritization as the design principles that specify how rank and weights structure the distribution of public goods. With this definition in mind, this paper provides a diagnostic analysis of the structural trade-offs that can necessarily emerge when adopting prioritization approaches. Specifically, our work asks the research question (RQ): \textbf{How does the structural design of public-sector prioritization algorithms shape resource allocation disparities?}

\begin{figure*}[]
\centering 
\includegraphics[scale=0.45]{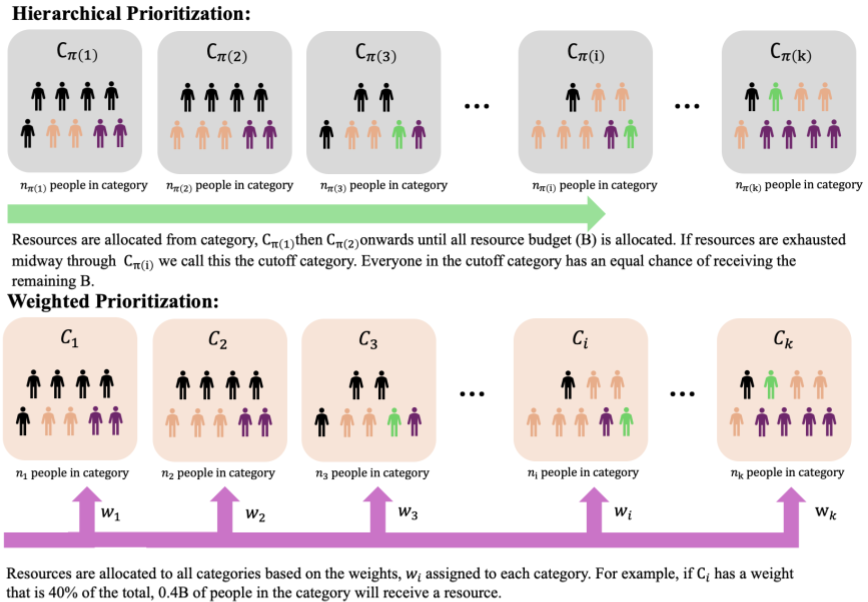}
\caption{Two Types of Prioritization Approaches for Resource Allocation: A strict hierarchical prioritization approach (1st row) and weighted prioritization approach (2nd row). Categories are composed of heterogeneous groups of individuals of diverse attributes}
\label{fig:prioritization_schemes}
\end{figure*}

To address our research question, we examine two prioritization approaches that underpin many public sector resource allocation systems, a hierarchical prioritization approach and weighted prioritization approach, which we depict in Figure \ref{fig:prioritization_schemes}. These two approaches are commonly applied to guide the prioritization and allocation of public resources, including in healthcare, housing in homelessness systems, and educational services \cite{eubanks2018automating, johnson22, Tracey_Garcia_2024_intermediation, kelly23}. We focus on the most basic version of each of these prioritization approaches as the aim of this work is to distill how the structural mechanisms underpinning these two  methods can systematically shape allocation outcomes. We make the following contributions to the paper: 

\begin{itemize}
    \item We frame public sector prioritization algorithms as policy interventions \cite{liu2025bridging} and examine the structural design of prioritization that underpins these models. We demonstrate how resource allocation disparities arise from inherent properties of these structures in Sections \ref{sec:sturcturalcomparisons} and \ref{sec:intersectionalimpact}. 
    

    \item We formalize how resource scarcity is a driver of inequality, particularly under a hierarchical prioritization regime in Section \ref{sec:explosive}. We demonstrate that under this regime, disparities in the likelihood of receiving resources (log-ratio differences in receiving resources) can become arbitrarily large as resources become scarce.

    \item We demonstrate how different evaluation metrics introduced in Section \ref{sec:metrics_measure} (absolute and log-ratio differences in resource allocation rates) for measuring the disparate impact of prioritization can reveal distinct experiential aspects of prioritization in Sections \ref{sec:explosive} and \ref{sec:dampen}. We present a naïve example of prioritization in homelessness to highlight these distinctions in Section \ref{sec:example}. 

\end{itemize}

In the following section, we detail the related works that have motivated this work.

\section{Related Works}

In recent years, often driven by neoliberal and market-oriented values, public agencies have increasingly sought to categorize and classify individuals across multiple dimensions to prioritize resources toward targeted groups, particularly in contexts of resource scarcity to promote allocation efficiency \cite{Dencik_Redden_Hintz_Warne_2019, Paul_Carmel_Cobbe_2024, eubanks2018automating, johnson22}. Public resources are not provided universally but instead, allocated through prioritization based on assessments of an individual’s or entity’s ‘deservingness’ or their ‘risk level’ \cite{johnson22}. Following the growing technical affordances of data-driven algorithmic systems, prioritization algorithms are being increasingly used to operationalize these assessments. Applications of these prioritization models in the public domain are far ranging: from criminal justice algorithms that predict areas at higher risk of criminal activity to guide the deployment of policing resources and homelessness algorithms that predict a person’s risk of experiencing chronic homelessness to prioritize the allocation of housing-related services \cite{moon24, ziosi24, showkat23, zilka22}. 

The algorithmic fairness community has extensively studied these systems, where a large body of literature has focused on model-centric evaluations such as improving model accuracy, validity, reliability, and fairness \cite{kleinberginherent, Chouldechova_2017, corbett-davies_2017, mashiat2022, coston23,jacobs20, impossibility21}. However, in response to studies that highlight the unintended harms associated with the deployment of these tools \cite{eubanks2018automating, saxena2021framework2,chapman22, pruss23}, Liu et al. \cite{liu2025bridging} argue for the need to move away from adopting a model-focused paradigm and instead, towards framing algorithmic systems as policy interventions that must be evaluated in relation to their implementation context. We respond to these calls and shift our attention to the very structural mechanisms that underpin public sector prioritization algorithms. We examine how structural prioritization mechanisms that govern the ranking and allocation of public goods interact with varying conditions of resource scarcity to produce disparate outcomes for impacted groups. We also demonstrate how different measures of disparity can reveal different experiential effects of prioritization in the public sector \cite{green20_falsepromise, liu2025bridging, selbst2019fairness}.

\section{Two Prioritization Approaches in the Public Sector}

In this paper, we examine the structural mechanisms of two prioritization approaches: hierarchical and weighted prioritization, depicted in Figure \ref{fig:prioritization_schemes}. Although both approaches partition individuals into groups and allocate resources based on a person’s group membership (i.e., members of certain groups are prioritized for resources), they represent two distinct and commonly observed approaches to distributing public goods \cite{stone97}. A hierarchical prioritization approach is a rank-based distribution approach that prioritizes resources to those with the highest need or deservingness \cite{stone97}. Emergency care triage systems, child welfare systems, and public school systems often adopt this approach to identify critically ill patients, children at risk of experiencing maltreatment, or students at risk of dropping out to direct scarce resources to them \cite{saxena2020human, kelly23, yancey23}. In hierarchical prioritization, individuals are often ranked based on individual-level measurements, and lower-ranked categories may be excluded altogether from receiving resources \cite{eubanks2018automating}. A weighted prioritization approach, on the other hand, is a group-based distribution strategy that often divides people based on immutable characteristics, such as ethnicity, race, or gender (often historical markers of privilege and disadvantage), and is employed when representation across groups is important \cite{stone97}. For example, this approach is applied in public housing and vaccine rollouts to ensure no groups are completely excluded from receiving resources \cite{plan_toronto, Persad_Peek_Emanuel_2020}. While these two prioritization approaches are not exhaustive, we introduce them to provide a diverse illustration of how different prioritization structures give rise to resource allocation disparities.


To formalize the mechanics of prioritization, we introduce the following set up. We consider a population of $N$ individuals in need of public resources, partitioned into $K$ disjoint categories, $\mathcal{C} = \{C_1, C_2,\dots,C_K\}$. These categories are determined by policymakers based on criteria such as risk scores, assessed level of need of a person, test scores, or other similar measures. Each category, $C_i$ contains $n_i$ individuals, where $\sum_{i=1}^K n_i = N$. We assume there is a fixed resource budget of $B$ resources, where a person can only receive one unit at most. Oftentimes in the public sector there is resource scarcity where $B < N$.

\subsection{Hierarchical Prioritization}
The first row of Figure \ref{fig:prioritization_schemes} depicts hierarchical prioritization. Under this scheme, categories are ordered according to a ranking criterion established by policymakers or an algorithm \cite{johnson22} and resources are allocated sequentially until the budget is exhausted. In cases where the remaining budget is insufficient to serve an entire category (we call this the "cutoff category"), we make the explicit assumption that everyone within that cutoff category is equally likely to receive the resource. We adopt this specific assumption as this allows us to isolate the impact of the hierarchical ranking structure on allocation outcomes (see Appendix \ref{sec:hierarchical_complicated} where we further discuss this assumption). While policymakers may assign identical ranks to multiple categories, we treat such instances as a single collapsed category for the purposes of this stylized model. When $n_i =1$ for all $K$ categories, this is the special case where we have individual-level prioritization.  

\begin{definition}[Hierarchical Prioritization] 
A hierarchical prioritization rule, $\pi$, ranks each category such that $C_{\pi(1)} \succ C_{\pi(2)} \succ \cdots \succ C_{\pi(K)}$. Given a resource budget $B$, resources are allocated to the categories in order of rank from highest to lowest. The probability $P_{\pi(i)}^{H}$ of an individual in category $\pi(i)$ receiving the resource is defined as: 

\begin{equation} \label{eq:hierarchical_prob}
P_{\pi(i)}^{H} =
\begin{cases}
1,
& \text{if } \sum_{j=1}^{i} n_{\pi(j)} \le B, \\

\dfrac{B - \sum_{j=1}^{i-1} n_{\pi(j)}}{n_{\pi(i)}},
& \text{if } \sum_{j=1}^{i-1} n_{\pi(j)} < B < \sum_{j=1}^{i} n_{\pi(j)}, \\

0,
& \text{if } \sum_{j=1}^{i-1} n_{\pi(j)} \ge B .
\end{cases}
\end{equation}
\end{definition}

Equation \ref{eq:hierarchical_prob} formalizes the sequential assignment of resources shown in the first row of Figure \ref{fig:prioritization_schemes}. In the Figure, the cutoff category for hierarchical prioritization is the category where the green arrow ends (i.e., $C_{\pi(i)}$). In this prioritization regime, individuals in categories ranked higher than the cutoff category (categories to the left of $C_{\pi(i)}$) are fully served and receive resources with a probability of $1$. When the budget, $B$ is partially exhausted in the cutoff category, individuals in this cutoff group receive resources with a probability equal to the remaining budget divided by the category size ($n_{\pi(i)}$). Any individuals in lower ranked categories (those in categories to the right of the cutoff category in the Figure) have a $0$ probability of receiving resources. 

\subsection{Weighted Prioritization}

The second row of Figure \ref{fig:prioritization_schemes} depicts weighted prioritization. This approach allocates resources across categories proportionally based on pre-assigned weights for each category. For example, if a category is assigned a weight that is 40\% of the total sum of weights, $0.4B$ resources will be assigned to the category (which means that $0.4B$ people in the category will receive resources). Although in resource scarcity, not every individual in a category will necessarily receive a resource, this approach ensures that multiple categories receive an allocation simultaneously, avoiding the "cutoff" effect seen in hierarchical prioritization where the budget may be exhausted before reaching lower-ranked groups. In the public sector, these weights, $w_i$, could represent the degree of need or deserving-ness for certain population groups (e.g., a fixed proportion of available homelessness shelters must go to Indigenous groups). 

\begin{definition}[Weighted Prioritization] \label{def:weightedprioritization}
A weighted prioritization rule assigns a positive weight, $w_i$ to each category, $C_i$. Given a resource budget $B$, the total amount of resources, $R_i$ allocated to category $i$ is \begin{equation}\label{eq:weighted_res} R_i = \frac{w_i}{\sum_{j=1}^{K} w_j} \cdot B \end{equation}. 

The probability $P_{i}^{W}$ of an individual in category $i$ receiving the resource is defined as: 

\begin{equation} \label{eq:weighted_prob} P_{i}^{W} = \min \left( 1, \frac{w_i \cdot B}{n_i \cdot \sum_{j=1}^{K} w_j} \right) \end{equation}

Note that as $B$ becomes large, the resource allocated to a specific category, $R_i$, can become greater than the size of the category, $n_i$. For this stylized example, we apply a “no reallocation after saturation” rule: we cap the probability at $1$, and any excess budget assigned to that category is considered unspent. This assumption allows us to isolate the effects of weighted prioritization in its most basic form. However, we acknowledge that in many real-world systems, excess resources are redirected to other categories, with reallocation mechanisms varying widely (E.g., weighted redistribution, uniform redistribution). 
\end{definition}

As illustrated by the purple arrows in the second row of Figure \ref{fig:prioritization_schemes}, every category receives a "slice" of the budget based on their assigned weight ($w_i$). Equation \ref{eq:weighted_prob} translates this category allocation weight into an individual probability. For example, if a category, $i$, of 10 people ($n_i=10$) is allocated 8 units ($R_i=8$), an individual has a $P^W_i=0.8$ probability of receiving resources. However, if this category instead has only 4 people ($n_i=4$), the "no reallocation after saturation" rule caps the probability at $P^W_i=1$.


\section{Structural Comparisons Between the Two Approaches} \label{sec:sturcturalcomparisons}

The structural design of the two prioritization approaches results in differences in how intra- and inter- group tensions arise within and between different categories, respectively.

\subsection{Intra-category Tensions}
Intra-category tensions arise when an individual's probability of receiving resources within a group decreases as the number of people in a category increases.  

In the \textbf{hierarchical approach}, as seen in Eq. \eqref{eq:hierarchical_prob}, for categories where the cumulative population is less than $B$, the probability of receiving resources for an individual, $P_{\pi(i)}^{H} =1$. However, for a cutoff category, $m$ where the remaining budget is insufficient to cover the entire group, the individual probability of a person receiving resources decreases as the group size grows. Formally, this decrease occurs at a decreasing rate:
\begin{equation} \label{eq:hierarchical_deriv} 
\frac{\partial P_{\pi(i)}^{H}}{\partial n_{\pi(i)}}= -\,\frac{B - \sum_{j=1}^{i-1} n_{\pi(j)}}{\bigl(n_{\pi(i)}\bigr)^2} < 0, \\
\frac{\partial^2 P_{\pi(i)}^{H}}{\partial n_{\pi(i)}^2}= \frac{2\left(B - \sum_{j=1}^{i-1} n_{\pi(j)}\right)}{(n_{\pi(i)})^3} > 0
\end{equation}

Following Equations \ref{eq:hierarchical_deriv}, we observe that for the cutoff category, adding a fixed number of people to the category reduces each person's probability of receiving resources more when that category is small than when it is large.

In the \textbf{weighted approach}, intra-category tensions are not confined within a single cutoff category as in the hierarchical approach but can be experienced across all categories. An increase in a category's group size results in a decrease in the individual probability of a person receiving resources at a decreasing rate for unsaturated categories: 
\begin{equation}
\label{eq:weighted_derivs} \frac{\partial P_{i}^{W}}{\partial n_{i}} = -\frac{w_i \cdot B}{n_i^2 \cdot \sum w_j} < 0, \quad \text{and} \quad \frac{\partial^2 P_{i}^{W}}{\partial n_{i}^2} = \frac{2 w_i B}{n_i^3 \sum w_j} > 0 \end{equation}

Here, the impact of a person's chances of receiving resources following a change in their category's group size is scaled by the weight assigned to their category. While high $w_i$ can favor the category in terms of total resources allocated, the marginal decrease in an individual's probability of receiving resources following an increase in category size is greater when the category is small than when it is large.

\subsection{Inter-category Tensions} Inter-category tensions arise when changes in one category's group size affect the likelihood that individuals in another category receive resources.

In the \textbf{hierarchical approach}, for any category $C_{\pi(k)}$ ranked lower than $C_{\pi(i)}$, an increase in the population of higher-ranked groups reduces the remaining resources for lower-ranked groups. In the marginal cutoff category, inter-category harm arises since:  
\begin{equation} \label{eq:inter_hier_neg} \frac{\partial P_{\pi(k)}^{H}}{\partial n_{\pi(i)}} = -\frac{1}{n_{\pi(k)}} < 0 \quad  \text{when } i < k\end{equation}

The derivative in Equation \ref{eq:inter_hier_neg} shows that adding an individual to a higher-ranked category marginally reduces each member’s probability of receiving resources in the cutoff category by $\frac{1}{n_{\pi(k)}}$. On the other hand, higher-ranked groups are unaffected by the population sizes of lower-ranked groups. Hierarchical prioritization approach thus creates a structural dynamic where low-priority groups are vulnerable to “budget exhaustion” caused by higher-priority groups. 

Under our definition of weighted prioritization, which assumes 1) weights assigned to each category are fixed constant and 2) that the “no reallocation after saturation” rule holds (Definition \ref{def:weightedprioritization}), there is no inter-category tension in the \textbf{weighted approach}, because changes in the size of one category do not affect the probability of resource receipt for other categories. However, we note that when either of the these assumptions is relaxed, inter-category tensions may arise as leftover resources can be redistributed to other categories. How these tensions emerge will depend on the excess resource redistribution policy. 

\section{Disparate Impact on Intersectional identities} \label{sec:intersectionalimpact}

In the previous section, we examined tensions that can arise within and between categories following hierarchical and weighted prioritization approaches. These tensions are simple, inherent structural properties of their respective allocation mechanisms. We now extend our analysis to examine how the design of hierarchical and weighted regimes can produce disparate impact for individuals with intersectional identities because classification of people into categories does not imply homogeneity within groups; instead, each category is often comprised of a heterogeneous population with diverse attributes. For example, in healthcare and child welfare systems, clients seeking services may be grouped based on risk level; however, within these groups, individuals may differ significantly by race, gender, or socioeconomic status \cite{eubanks2018automating, Benjamin_2019, angwin, cheng2024algorithmassisteddecisionmakingracial}. Prior works has shown how such categorized groups can systematically marginalize specific populations, particularly when these categories are strongly correlated with protected attributes such as poverty or race \cite{gerchick_23, obermeyer19, eubanks2018automating, Cronley_2022}. 

\subsection{Metrics to measure disparate impact} \label{sec:metrics_measure}

To formalize disparate impacts of prioritization on subgroups within categories, we consider a set of population subgroups: 
\[ \mathcal{S} = \{s_1, s_2, \dots, s_m\} \] distributed across the $K$ categories. In Figure \ref{fig:prioritization_schemes}, these subgroups are depicted by different colored figures depicted within categories. Let $n_{i,s}$ denote the number of people in $C_i$ who belong to subgroup, $s$. The total number of people in category $n_i = \sum_{s \in \mathcal{S}} n_{i,s}$ and the total population of a specific subgroup is $N_s = \sum_{i=1}^K n_{i,s}$. Within any category $C_i$, we assume all individuals have an equal probability $P_i$ of receiving a resource regardless of their subgroup membership. We intentionally assume no within-category discrimination to isolate structural effects and avoid introducing within-category bias. 

To understand the disparate impacts of prioritization, we compare how the proportion of the total subgroup population expected to receive a resource differs between subgroups and across varying budgetary conditions. Below, we define the metrics we use to quantify subgroup allocation disparities.

\begin{definition}[Subgroup Resource Receipt Rate] The expected resource receipt rate for subgroup $s$, denoted $G_s(B)$, is the proportion of the total subgroup population that is expected to receive a resource:

\begin{equation} \label{eq:group_receipt_rate} G_s(B) = \frac{\sum_{i=1}^{K} n_{i,s} \cdot P_{i}(B)}{N_s} \end{equation} 

where $P_i(B)$ represents the probability of an individual in category $C_i$  receiving a resource given a total budget $B$.
\end{definition}

\textbf{The Subgroup Resource Receipt Rate} represents the expected proportion of people in a particular subgroup who will receive resources given a resource budget of $B$. For example, if half of the people in subgroup $s_1$ are in category 1 with a 100\% chance of receiving resources, and the other half are in category 2 with a 20\% chance of receiving resources, then $0.6$ of those in subgroup, $s_1$, are expected to receive a resource.


To understand disparities in resource allocation between subgroups, we compare Subgroup Resource Receipt Rates across two metrics: log-ratio 
differences, $\ln(RD(B))$, and absolute difference, $AD(B)$, which we define below.

\begin{definition}[Absolute Difference in Subgroup Resource Receipt Rate] The absolute difference between two expected subgroup resource receipt rates (e.g., between subgroups $s_1$ and $s_2$) at budget $B$ is defined as: \begin{equation} \label{eq:ad}AD(B) = \left|G_{s_1}(B) - G_{s_2}(B)\right| \end{equation} \end{definition}  

\textbf{The Absolute Difference ($AD$)} captures how unequally resources are allocated between two subgroups under a given prioritization regime. For example, if 60\% of individuals in subgroup, $s_1$ and 40\% of people in subgroup, $s_2$ are expected to received resources then $AD(B)=0.2$.

\begin{definition}[Log-Ratio Difference in Subgroup Resource Receipt Rate]\label{def:ratiodifferencein} The ratio difference between two expected subgroup resource receipt rates at budget $B$ is: \begin{equation} \label{eq:rd}RD(B) = \frac{G_{s_1}(B)+ \epsilon}{G_{s_2}(B)+ \epsilon}\end{equation}
where $0<\epsilon\ll1$ is a small smoothing constant to ensure the ratio is well defined because $RD(B)$ becomes undefined when denominator subgroup resource receipt rates equals $0$. Taking the log of $RD(B)$, we define the log-ratio difference in subgroup resource receipt rates as: 
    \begin{equation} \label{eq:logrd}  \ln(RD(B)) = \ln\left(\frac{G_{s_1}(B) + \epsilon}{G_{s_2}(B) + \epsilon}\right) \end{equation}

This measure provides a useful metric to understand differences in magnitude between the two groups' receipt rates. We note the absolute value of this log-ratio also provides a useful symmetric measure of relative disparity between subgroups. E.g., Taking the absolute value of $\ln(RD(B))$ allows us to treat the ratio difference of $RD(B)=2$ the same as a ratio difference of $RD(B)=\frac{1}{2}$. Additionally, we note that the absolute value of the derivative of this Log-Ratio further provides a useful metric to capture the sensitivity of the prioritization scheme to budgetary fluctuations as this metric is also symmetric (explained further in Section \ref{sec:explosive} and \ref{sec:dampen}).
\end{definition}


We apply a small epsilon in Definition \ref{def:ratiodifferencein} to ensure these quantities are well defined when the denominator subgroup resource receipt rate equals zero. In these instances, the resulting values should not be interpreted as an empirically meaningful measure of disparity. This differs from cases where the subgroup resource receipt rate is very low and near zero; in this instance disparity values reflect empirically substantive differences in expected resource allocation, which we discuss further below.

The \textbf{Log-Ratio Difference in Subgroup Receipt Rate} captures the proportional disparity of receiving resources between two subgroups. While the Absolute Difference measures the size of the gap between two Subgroup Receipt Rates, the Log-Ratio Difference focuses on how many times more likely one subgroup is to receive resources over another. For example, if subgroup $s_1$ has an expected Resource Receipt Rate of 0.8 and subgroup $s_2$ has a rate of 0.4, then $RD(B)=2$, meaning those in subgroup $s_1$ are twice as likely to receive resources as those in subgroup $s_2$. The logarithmic transformation (Log-Ratio Difference in Subgroup Resource Receipt Rate) provides mathematical symmetry, ensuring that an advantage for one subgroup is given the same weight as an equivalent advantage for the other.  For instance, if $s_1$ is twice as likely as $s_2$ to receive resources, the Log-Ratio is approximately $0.69$; if the situation were reversed and $s_2$ were twice as likely, the value would be $-0.69$. Taking the absolute value of this metric allows us to measure the intensity of the disparity regardless of which subgroup is more advantaged.






\subsection{Hierarchical Prioritization can Produce Explosive Subgroup Disparities in High Resource Scarcity} \label{sec:explosive}

Beyond the inter- and intra-category tensions discussed in Section~\ref{sec:sturcturalcomparisons}, hierarchical prioritization can give rise to intersectional disparities in resource receipt when categories differ in their composition of intersectional subgroups. When the resource budget, $B$, falls within the cutoff category, $C_{\pi(m)}$ (i.e., the budget cannot serve all the individuals in the cutoff category), the marginal benefit to subgroup $s$ as the budget, $B$ changes is as follows (see Appendix \ref{sec:derivation_G_s} for the derivation): 
\begin{equation}\label{eq:derivative_G_s}
\frac{dG^H_s}{dB} = \frac{1}{N_s} \cdot \frac{n_{\pi(m), s}}{n_{\pi(m)}}
\end{equation}

Equation \ref{eq:derivative_G_s} demonstrates that when the budget increases, a subgroup's chances of receiving resources depend on the subgroup's density within the cutoff category and the subgroup's size across all categories. In short, if the subgroup is heavily concentrated in the cutoff category, a unit increase in the resource budget will substantially increase the subgroup's Resource Receipt Rate. However, if the subgroup population is spread thin across many categories, a unit increase in the budget will only yield a modest increase in the subgroup's Resource Receipt Rate.


We observe the most interesting findings when we compare the Absolute Difference in Resource Receipt Rates, $AD^H(B)$ and Log-Ratio Differences in Resource Receipts rates, $\ln(RD^H(B))$ as they respond to changes in $B$. In hierarchical prioritization, $AD^H(B)$ moves in a piecewise linear fashion as we show below:

\begin{equation}\label{eq:derivative_AD}
    \frac{d}{dB} AD^H(B) = \text{sgn}(G^H_{s1} - G^H_{s2}) \left( \frac{n_{\pi(m),s_1}}{N_{s_1} \cdot n_{\pi(m)}} - \frac{n_{\pi(m),s_2}}{N_{s_2} \cdot n_{\pi(m)}} \right)
\end{equation}

The derivative in Equation \ref{eq:derivative_AD} is undefined when $G^H_{s1} = G^H_{s2}$ and its sign flips as the budget crosses this point. As the budget increases to cover more categories, $AD^H(B)$ changes at a constant rate within each category. The rate is dependent on the difference in the subgroups' proportional representation within the cutoff category relative to each subgroup's total size. For example, if a subgroup $s_1$ has a higher receipt rate than subgroup $s_2$ (I.e., $G^H_{s1} > G^H_{s2}$) and there is a higher proportion of $s_1$ in the cutoff category compared to $s_2$, then we would expect to see the gap between the Resource Receipt Rates for the two subgroups to increase as the budget increases. Conversely, if a subgroup $s_2$ has a higher receipt rate than subgroup $s_1$ (I.e., $G^H_{s1} <G^H_{s2}$) and there are a higher proportion of $s_1$ in the cutoff category compared to $s_2$, then we would expect to see the gap between the Resource Receipt Rates for the two subgroups to decrease as the budget increases.


While Equation \ref{eq:derivative_AD} shows that $AD^H(B)$ moves in a piecewise linear fashion, $RD^H(B)$ exhibits greater volatility and can create explosive disparities between subgroups in hierarchical approaches. The sensitivity of the $RD^H(B)$ is best captured by observing the absolute value of the derivative of the log-ratio differences (see Definition \ref{def:ratiodifferencein}):
\begin{equation}\label{eq:derivative_RD}
\left|\frac{d}{dB} \ln(RD^H(B))\right| 
= \left|\frac{G'^H_{s1}(B)}{G^H_{s1}(B)+\epsilon} - \frac{G'^H_{s2}(B)}{G^H_{s2}(B)+\epsilon} \right| \\
= \left|\frac{\frac{n_{\pi(m),s_1}}{N_{s_1} \cdot n_{\pi(m)}}}{G^H_{s1}(B)+\epsilon} - \frac{\frac{n_{\pi(m),s_2}}{N_{s_2} \cdot n_{\pi(m)}}}{G^H_{s2}(B)+\epsilon}\right| 
\end{equation}

As noted in Definition \ref{def:ratiodifferencein}, the absolute value of the derivative of $\ln(RD^H(B))$ captures how sensitive the relative likelihood of receiving resources is for one subgroup compared to another as the resource budget changes. Equation \ref{eq:derivative_RD} shows that the derivative is calculated by the inverse of the subgroup receipt rates. Our findings imply that when 1) the resource budget falls within the cutoff category, 2) subgroup resource receipt rates that are being compared are positive, and 3) when there is high resource scarcity such that a subgroup's resource receipt rate is very low and near zero, then even slight budgetary shifts can create explosive changes in $RD^H(B)$. 

In short, $AD^H(B)$ changes steadily as the budget grows but $\ln(RD^H(B))$ can fluctuate sharply. This happens because when a subgroup receives very few resources, even a small budget increase can dramatically change their relative chances of getting resources. For example, consider the scenario where $G^H_{s1}(B)=0.4$ and $G^H_{s2}(B)=0.01$. Here the Absolute Difference is $AD^H(B)=0.39$ while the Log-Ratio Difference is $\ln(RD^H(B))=3.69$. Now, if a small budget increase adds just 0.01 to each rate ($G^H_{s1}(B)=0.41$ and $G^H_{s2}(B)=0.02$), the Absolute Difference will remain at $0.39$, but the Log-Ratio Difference drops sharply to $\ln(RD^H(B))=3.02$. The divergence between $AD^H(B)$ and $\ln(RD^H(B))$ as the budget changes is important because for impacted individuals who are hoping to receive resources, $\ln(RD^H(B))$ may be a more pertinent measure that reflects their experiential relative likelihood of receiving resources.


\subsection{Weighted Prioritization Avoids Explosive Subgroup Disparities } \label{sec:dampen}

We now examine the weighted prioritization approach. By assigning a nonzero weight ($w_i>0$) to all categories, this approach provides a way to ensure that no category is entirely excluded from receiving resources. In weighted prioritization, as the resource budget grows, the resource share allocated to each category will increase. Eventually, the budget will become large enough to allocate resources to every individual within a specific category ($n_i$), reaching what we define as the saturation threshold. Because each category may have a different size and weight, the saturation thresholds will occur at different budgetary thresholds, E.g., $\{B_{sat,1}, B_{sat,2} \dots B_{sat,k}\}$. In the following paragraphs, we denote $\mathcal{A}(B)$ as the set of categories that are not yet saturated (i.e., $R_i<n_i$) and $\mathcal{F}(B)$ as the set of saturated categories (i.e., $R_i \geq n_i$), the group receipt rate for subgroup $s$ is (note for our stylized model, we assume weights are static):


\begin{equation}\label{eq:weighted_gr_receipt_rate}
G_s^W(B)
=
\frac{1}{N_s}
\left(
\sum_{i \in \mathcal{F}(B)} n_{i,s}
\;+\;
\sum_{i \in \mathcal{A}(B)} B
\left[
\frac{w_i \cdot n_{i,s}}{n_i \cdot \sum_{j=1}^K w_j}
\right]
\right)
\end{equation}

Equation \ref{eq:weighted_gr_receipt_rate} shows how the overall Resource Receipt Rate for a subgroup is determined under Weighted Prioritization. We first count all subgroup members in the saturated categories ($\mathcal{F}(B)$), where every individual in these categories is served a resource. To this, we add the expected number of subgroup members who receive resources in the unsaturated categories ($\mathcal{A}(B)$). Finally, we divide this total by the subgroup’s population size ($N_s$).


Similar to hierarchical prioritization (Equation \ref{eq:derivative_AD}), the absolute difference in resource receipt rates, $AD^W(B)$ moves in a piecewise fashion as the budget changes:

\begin{equation}\label{eq:weighted_ad_derivative}
\frac{d}{dB} AD^W(B) = \text{sgn}\left(G_{s_1}^W(B) - G_{s_2}^W(B)\right) \left[ \sum_{i \in \mathcal{A}(B)} \frac{w_i}{\sum_{j=1}^K w_j} \left( \frac{n_{i,s_1}}{n_i N_{s_1}} - \frac{n_{i,s_2}}{n_i N_{s_2}} \right) \right]
\end{equation}

Equation \ref{eq:weighted_ad_derivative} shows that the change in the Absolute Difference in Resource Receipt Rate as the budget increases depends on categories that are not yet saturated. Each unsaturated category can either increase or decrease the gap, depending on which subgroup is more strongly represented in that category relative to the subgroup's total population. The magnitude of this effect is scaled by the weight assigned to the category; a higher weight means that the category has a larger influence on whether $AD^W(B)$ widens or narrows.

The greatest difference between the weighted and hierarchical approaches can be observed in the Log-Ratio Difference of Resource Receipt Rates, $\ln(RD^W(B)$). When subgroups being compared have positive resource receipt rates, we no longer observe the explosive increase in log-ratio differences we observed in hierarchical prioritization. In weighted prioritization, prior to any category becoming saturated, the Log-Ratio Difference in Subgroup Receipt Rates, $\ln(RD^W(B))$ is:

\begin{equation}\label{eq:ln_RDW_deriv} \ln(RD^W(B)) = \ln(\frac{G_{s_1}(B)+\epsilon}{G_{s_2}(B)+\epsilon}) = \ln\left(\frac{ \frac{B}{N_{s_1} \sum w_j} \sum_{i \in \mathcal{A}(B)} \frac{w_i n_{i,s_1}}{n_i} + \epsilon}{ \frac{B}{N_{s_2} \sum w_j} \sum_{i \in \mathcal{A}(B)} \frac{w_i n_{i,s_2}}{n_i} + \epsilon}\right)\end{equation}


When the budget is larger relative to the smoothing term $\epsilon$ then $B$ approximately cancels out, making the Log-Ratio Difference nearly constant. This implies that, as long as no category becomes saturated, the relative likelihood of receiving resources for one subgroup compared to another does not fluctuate greatly as the budget increases. As the budget continues to increase, and once one or more categories become saturated, then $RD(B)\to1$ (equivalently $\ln(RD(B))\to0$) (reflecting that the subgroups are moving towards parity). These findings hold under the "no reallocation after saturation" rule. If this rule is no longer upheld and excess resources can be allocated to other categories, then we could observe lower absolute and log-ratio disparities (however, this is contingent on how excess resources are redistributed).


\subsection{Visualizing Resource Allocation Disparities between Subgroups: An Example in Homelessness} \label{sec:example}

In this section, we visualize the disparate impact of prioritization in the context of homelessness where there is high demand across North American cities for emergency shelter beds and affordable permanent housing. In response to rising demand for housing-related services, federal governments in North America have adopted prioritization approaches to allocate housing-related resources, including emergency shelter spaces \cite{Ecker2022, reachinghomes}. 

\begin{figure*}[]
\centering 
\includegraphics[scale=0.35]{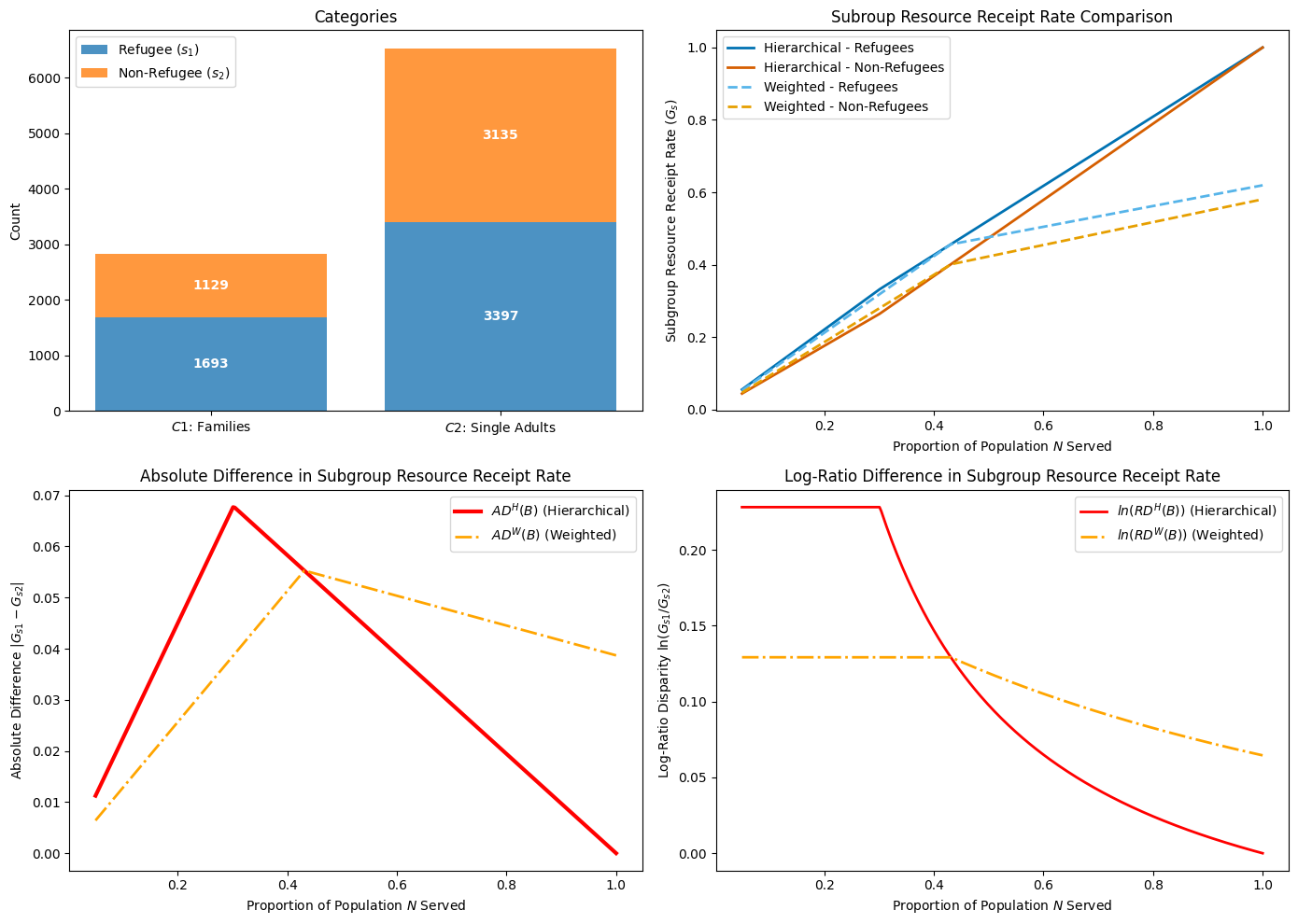}
\caption{Comparing Subgroup Resource Receipt Rates between Two Unhoused Groups, Families and Single Adults for Weighted and Hierarchical Prioritization: (Top-Left) Population distribution across categories. (Top-Right) Subgroup resource receipt rate ($G_s$) comparisons as a function of total population served. (Bottom-Left) Absolute Difference ($AD(B)$) in subgroup resource receipt rates between refugees and non-refugees. (Bottom-Right) Log-Ratio differences ($\ln(RD(B))$) in subgroup resource receipt rates between refugees and non-refugees}
\label{fig:AD_RD}
\end{figure*}

We pull publicly available shelter population data from a large North American city \cite{streetneeds24, sheltersystemflow_realtime} collected in October 2024 to obtain counts of families and single adults where this data can be disaggregated by their refugee claimant status. Ethnographic scholarship suggests that families are, in the most general sense, perceived as more vulnerable than single adults and are often prioritized for services \cite{passaro14}. As a result, in this example, we consider two regimes that prioritize families over single adults for services: (1) a hierarchical prioritization regime where families are prioritized over single adults for access to permanent housing. And (2) a weighted prioritization regime where we arbitrarily assign weights of 0.7 and 0.3 to Families ($C_1$) and Single Adults ($C_2$), respectively. The distribution of refugee and non-refugee subgroups within these categories is shown in the top-left panel of Figure \ref{fig:AD_RD}. 

As seen in the top-right panel of Figure \ref{fig:AD_RD}, regardless of the prioritization approach, refugees have a higher Subgroup Resource Receipt Rate ($G_s$) than non-refugees. Because resources are allocated sequentially under hierarchical prioritization (solid lines), $G^H_s$ approaches $1$ as $B \to N$ (i.e., when the budget serves the entire $N$ population). However, because we apply static weights and the "no reallocation after saturation rule" in the weighted prioritization approach (dashed lines), as the assignment of resources to a category reaches saturation (i.e., $R_i=n_i$), $G^W_s$ levels off and does not reach $1$ as $B \to N$ since there are unused resources.

The bottom left panel illustrates how the absolute difference in receipt rates for the subgroups varies with changes in resource availability. Regardless of the prioritization approach, we observe that $AD(B)$ moves in a piecewise linear manner and is bounded. Furthermore, as $B$ increases from 0, the budget is being allocated to the subgroup that holds a higher allocation probability in this regime (in our example, refugees), increasing $AD(B)$. However, as the budget begins to be allocated to the subgroup with the lower allocation probability in our regime (in our example, non-refugees), $AD(B)$ begins to decrease. Again, due to static weights and the "no reallocation after saturation rule," in weighted prioritization, $AD^W(B)$ never reaches $0$ at full resource availability.

In the bottom right panel, we see more volatility in $\ln(RD^H(B))$ in hierarchical prioritization compared to $AD(B)$. In high resource scarcity, the magnitude of the receipt rate disparity is at its peak; and $\ln(RD^H(B))$ remains flat when resources are allocated only to the first category. On the other hand, $\ln(RD^W(B))$ remains flat for a longer budget interval as resources are distributed across multiple categories simultaneously. $\ln(RD^W(B))$ only begins to curve downwards the moment the higher weighted category, Families ($C_1$), are all assigned resources. 

Our study's most compelling findings emerge when we compare $\ln(RD^H(B))$ and $AD^H(B)$ under hierarchical prioritization. While $AD^H(B)$ remains bounded and relatively low across all budget levels, $\ln(RD^H(B))$ reveals severe disparities when there is high resource scarcity. From a policymaker’s  perspective, the relatively stable, low values in $AD^H(B)$ might suggest that subgroup disparities are acceptable. However, from the lived perspective of an unhoused person, the impact of prioritization could be experienced through $\ln(RD^H(B))$. In our example, at low budget levels, the relative likelihood of receiving housing is much lower for non-refugees. Consider when only 5\% of the shelter population will receive housing (a realistic assumption for the City \cite{sheltersystemflow_realtime}). The resource receipt rate for a non-refugee, $G_{Nonrefugee}(B=468)=0.044$ while for a refugee, $G_{Refugee}(B=468)=0.055$. In absolute nominal terms, the differences are small. However, from the perspective of a non-refugee seeking housing, refugees would appear to have 26\% higher chance of obtaining housing. In contrast, by distributing resources across categories, the weighted prioritization can dampen this disparity to 13\%. Overall, our example suggests that outcomes for all subgroups worsens with increasing housing scarcity. However, individuals in non-prioritized subgroups may feel a greater comparative disadvantage.

\section{Discussion}

In this work, we examined how the structural design of two common prioritization approaches applied in the public sector (weighted and hierarchical) can lead to resource allocation disparities between and within groups (Section \ref{sec:sturcturalcomparisons}). We further demonstrate how subgroup allocation disparities can arise from the inherent properties of the prioritization regime itself (Section \ref{sec:intersectionalimpact}). Our findings from Section \ref{sec:explosive} in particular highlight the implications of applying hierarchical prioritization to allocate public goods under conditions of extreme budget scarcity. In these contexts, we observe that resource allocation disparities become acutely sensitive to the proportional representation of subgroup population sizes. Although these effects can be mitigated under weighted prioritization (Section \ref{sec:dampen}), overall, we observe relative allocation disparities between population groups arise from the structural mechanisms of prioritization. Our findings are important because we show distributive disparities emerge as direct consequences of intentional and normative policy choices on who to prioritize or not.

The distributive disparities identified in our work carry important social and policy implications for the allocation of scarce public resources using prioritization algorithms. Two key implications arise from our work: 1) we highlight the often overlooked trade-offs between efficiency and equity-related concerns in popular policy discourse on algorithmic tools, and 2) how allocation disparities may be experienced differently depending on whether they are viewed from a policymaker’s perspective or from the impacted individual. Predictive risk prioritization algorithms (a variant of hierarchical prioritization) have long been touted for their ability to target and deliver social services efficiently \cite{Dencik_Redden_Hintz_Warne_2019}. In recent years, amidst the growing technical affordances of AI tools, there is renewed interest in leveraging such computational models to deliver greater governmental efficiencies \cite{cohere, dhs_genai, govt_uk_ai, moon26_promises}. Oftentimes, governmental "efficiency" is conflated with "doing more with less," as witnessed by the cuts made under the US Department of Government Efficiency (DOGE) in 2025 \cite{whitehouse_dodge}. Our findings demonstrate that achieving these efficiency gains can invoke critical normative trade-offs. As a result, efficient prioritization of resources should not be necessarily conflated with the reduction of resource availability as this can come at the cost of stark allocation disparities when resources are scarce.


Stone \cite{stone97} argues that distribution processes can strengthen or undermine social cohesion. Our work does not seek to de-legitimize prioritization but rather to clarify the tradeoffs prioritization structures create under scarcity. We hope our work encourages renewed scrutiny of the role that algorithmic prioritization tools should play in public sector resource allocation tasks and their societal impact. Prioritization algorithms primarily address the question of “who” should receive scarce public goods. In this work, we make explicit how different prioritization structures give rise to disparate impacts, and how these impacts depend on the perspective from which they are evaluated. We introduce and compare two metrics, the absolute difference in subgroup receipt rates ($AD(B)$) and the log-ratio difference in subgroup receipt rates ($\ln(RD(B)$)), to demonstrate how the prioritization regime is experienced/perceived depends on the lens through which it is measured. From a top-down, social planner's perspective, policymakers may deem a prioritization regime acceptable even under high resource scarcity due to small nominal differences in $AD(B)$. However, from the perspective of impacted individuals, ratio comparisons ($\ln(RD(B))$) may be a more appropriate metric as individuals experience resource allocation as competition within their own subgroup and assess their relative likelihood of receiving resources in comparison to other subgroups.

The approaches we scrutinize contrast with other non-algorithmic allocation strategies implemented in the public sector, such as lotteries or first-come-first-served (FCFS) used in the provision of resources, for example, the provision of admission spots in charter schools \cite{Cohodes_Roy_2025}. These approaches can arguably offer a stronger form of procedural fairness around “how” scarce goods should be allocated \cite{stone97}. And yet, when resources are scarce and when some individuals face acutely higher needs over others, adopting a non-prioritarian allocation approach may be deemed ethically or politically infeasible \cite{walzer_spheres}.

Returning to our example in Section \ref{sec:example}, which visualizes the disparate impact of prioritization under varying resource availability, we observe that our findings succinctly capture the experiential disparities documented in ethnographic studies on how different population groups experience homelessness in North America. Previously, Passaro \cite{passaro14} argued that, shaped by normative societal biases regarding the worth and deservingness of different groups, North American single adult men have been placed at the bottom of every rung in receiving housing-related support. Passaro claimed that policies that “ignore men are helping to create increasingly desperate and violent men. This is not the result that anybody had in mind in attempting to protect women and children; it is the unintended outcome of what are well-meaning efforts to help.” Although homelessness systems have undergone substantial changes since Passaro’s writing, particularly following the adoption of centralized prioritization approaches (often referred to as coordinated entry and access systems \cite{Ecker2022}), the paradoxical effects of prioritization remain. Through our work, we demonstrate that adopting prioritization under high resource scarcity will inevitably lead to the creation of worse-off groups, and the magnitude of disparity between groups is directly tied to the structure of prioritization itself.

\section{Limitations and Future Work}
This work examines how disparities can arise from the structural design of two prioritization approaches in their most basic form by making several simplifying assumptions (e.g., assuming static weights and a “no reallocation after saturation” rule in weighted prioritization, and equal chances of resource receipt for all individuals within the cutoff category in hierarchical prioritization). Furthermore, we focused on one round of resource allocation through prioritization and did not consider how disparities can emerge under other distribution policies, such as first-come-first-serve or random lotteries. In practice, public-sector prioritization algorithms operate in different policy and organizational contexts, where multiple distributive policies and algorithms operate simultaneously \cite{Tracey_Garcia_2024_intermediation, saxena2021framework2}. These factors may amplify or diminish the disparities we identify in our paper. For example, Saxena and Guha \cite{saxena_guha24_jrc} show that once an allegation of child abuse is made and a family engages with the child welfare system, multiple prioritization algorithms may be deployed at multiple stages of a family’s journey in the system. Thus, the extent to which the disparities we observe in our work manifest in practice will depend on specific domain implementation contexts and the discretionary decision-making of street-level bureaucrats \cite{feng22, moon_datafication, kawakami2022, karusala_contestability24}. Lastly, while this work focused on the structural mechanics of “who” gets resources in prioritization algorithms, future work can examine how disparities also emerge when algorithms determine “what” type of resources are allocated and “how” they are delivered to individuals.


\section{Conclusion}
In this study, we examined how the structural design of prioritization (weighted and hierarchical prioritization) commonly applied in public sector prioritization algorithms give rise to disparate allocation outcomes for different population groups of intersectional identities. We introduced two evaluation metrics to assess the disparate impact of prioritization (log-ratio differences and absolute differences) and highlight how the metrics reveal different perspectives on how a prioritization policy is experienced and perceived. 

\section{Generative AI Usage Statement}
Generative AI (Gemini and Chatgpt) was used to check the text for typographical errors and assist in the formatting of Figure \ref{fig:AD_RD}. The authors reviewed all these suggestions generated by the AI models.

\begin{acks}
This research was supported by the NSERC Discovery Early Career Researcher Grant RGPIN-2022-04570 and University of Toronto School of Cities Catalyst Grant. Opinions, findings, and conclusions expressed in this paper are those of the authors. We sincerely thank our anonymous reviewers whose suggestions and comments helped improve this manuscript.
\end{acks}

\bibliographystyle{ACM-Reference-Format}
\bibliography{sample-base}


\appendix
\clearpage
\onecolumn

\section{Impact of relaxing equal resource allocation chances in the cutoff category in hierarchical prioritization}
\label{sec:hierarchical_complicated}

When we relax the assumption that everyone within the cutoff category should have an equal chance of receiving resources, within-category discrimination will also contribute to resource allocation disparities, further complicating the structural effects of hierarchical prioritization. If this condition is relaxed to allow for example, within-category weighting of subgroups $\mathcal{S}$ then $P^H_{\pi(i),s}$ (probability of an individual in category $\pi(i)$ and subgroup $s$ receiving the resource) would also be impacted by the weights assigned to the subgroup $s$.


\section{Derivation for Equation \ref{eq:derivative_G_s}} \label{sec:derivation_G_s}

The expected number of people with attribute $s$ who will receive the resource within the group $i$ is:

    \[
\mathbb{E}\!\left[ R_{\pi(i),\,s} \right]  =
\begin{cases}
n_{\pi(i),s},
& \text{if } \sum_{j=1}^{i} n_{\pi(j)} \le B, \\

n_{\pi(i),s} \left( \frac{B - \sum_{j=1}^{i-1} n_{\pi(j)}}{n_{\pi(i)}} \right),
& \text{if } \sum_{j=1}^{i-1} n_{\pi(j)} < B < \sum_{j=1}^{i} n_{\pi(j)}, \\

0,
& \text{if } \sum_{j=1}^{i-1} n_{\pi(j)} \ge B .
\end{cases}
\]

Then, $G^H_s$ is as follows where $m$ depicts the category that marginally receives resources: 

\[
G^H_s = \frac{\mathbb{E}[R_s]}{N_s}
= \frac{\sum_{i=1}^{K} n_{\pi(i),s} \cdot P_{\pi(i)}}{\sum_{i=1}^{K} n_{\pi(i),s}} 
= \frac{1}{N_s} \left[ \sum_{i=1}^{m-1} n_{\pi(i),s} + n_{\pi(m),s} \left( \frac{B - \sum_{j=1}^{m-1} n_{\pi(j)}}{n_{\pi(m)}} \right) \right]
\]

Now you can differentiate the above expression. 

\section{Derivation for Equation \ref{eq:derivative_AD}} \label{sec:derivation_deriv_AD}
$\frac{d}{dB} AD^H(B)$ can be found by differentiating the below expression: 
\[
AD^H(B) = \left| \frac{\sum_{i=1}^{m-1} n_{\pi(i),s1} + \frac{n_{\pi(m),s_1}}{n_{\pi(m)}}(B - \sum_{i=1}^{m-1} n_{\pi(i)})}{N_{s_1}} - \frac{\sum_{i=1}^{m-1} n_{\pi(i),s2} + \frac{n_{\pi(m),s_2}}{n_{\pi(m)}}(B - \sum_{i=1}^{m-1} n_{\pi(i)})}{N_{s_2}} \right|
\]

\section{Derivation for Equation \ref{eq:derivative_RD}} \label{sec:derivation_deriv_RD}

Note for cutoff category $m$,
\[
RD^H(B) = \frac{G^H_{s_1}(B)+\epsilon}{G^H_{s_2}(B)+\epsilon} = \frac{\frac{1}{N_{s1}}(\sum_{i=1}^{m-1} n_{\pi(i),s1} + \frac{n_{\pi(m),s_1}}{n_{\pi(m)}}(B - \sum_{i=1}^{m-1} n_{\pi(i)}))+\epsilon}{\frac{1}{N_{s2}}(\sum_{i=1}^{m-1} n_{\pi(i),s2} + \frac{n_{\pi(m),s_2}}{n_{\pi(m)}}(B - \sum_{i=1}^{m-1} n_{\pi(i)}))+\epsilon} 
\]

Note that with a very low budget, only the first category is served and $RD^H(B) = \frac{N_{s_2}}{N_{s_1}} \cdot \frac{n_{\pi(1),s_1}}{n_{\pi(1),s_2}}$ when we assume $G^H_{s2}(B) >0$. As $B \to N$, then $RD^H(B) \to 1$.

\end{document}